\documentclass[aps,twocolumn,pra,showpacs,floatfix]{revtex4}
\usepackage{epsfig}
\usepackage{graphicx}
\usepackage{dcolumn}

\begin{document}

\title{Atomic Ionization by 
 keV-scale Pseudoscalar
Dark Matter Particles}
\author{V. A. Dzuba and V. V. Flambaum}
\affiliation{School of Physics, University of New South Wales, Sydney 2052, 
Australia}
\author{M. Pospelov}
\affiliation{Department of Physics and Astronomy, University of Victoria, 
Victoria, British Colombia, V8P IAI, Canada}
\affiliation{Perimeter Institute for Theoretical Physics, Waterloo,
Ontario, N2J 2W9, Canada}

\date{\today}

\begin{abstract}

Using the relativistic Hartree-Fock approximation, we calculate 
the rates of atomic ionization by absorption of pseudoscalar particles  
in the mass range from 10 to $\sim$ 50 keV.
We present numerical results for atoms relevant for the direct dark
matter searches ({\em e.g.} Ar, Ge, I and Xe), 
as well as the analytical formula which fits numerical calculations with 
few per cent accuracy and may be used for  multi-electron atoms,
 molecules and condensed matter systems.
\end{abstract}


% 14.70.Pw - Other gauge bosons 
% 95.35.+d - Dark matter
% 32.80.Fb - Photoionization of atoms and ions

\maketitle

\section{Introduction}

Different lines of evidence consistently point to the existence of cold dark
 matter,
that comprises about 1/5 of the current energy budget of the Universe, 
%!!!
5 times more than the usual matter (see, {\em e.g.} reviews
\cite{Jungman,Bernone,Steffen,Komatsu,Carosi}). 
So far, the dark matter is seen only via its gravitational 
effects, which offers few clues of its true identity. Stable on cosmological 
time scales elementary 
particles with weak or super-weak interactions to visible matter 
naturally arise in many extensions of the Standard Model 
(see {\em e.g.} \cite{Jungman,Steffen}),
and may comprise the bulk of dark matter. 
The last two decades have seen a many-fold expansion 
of the experimental program aimed at direct detection of energy 
 deposited by the dark matter particles in the collision with atoms in 
 low radiation environments. So far no positive dark matter signal has 
 been detected, which places important constraints on many extensions of
 the Standard Model that predict such particles. 
 
Heavy stable particles (WIMPs) with masses above $1~{\rm GeV}$ can
manifest themselves by depositing some fraction of their kinetic
energy in the keV-to-100 keV range in the process of scattering with
atoms. Many dark matter search strategies are tuned with such
signature in mind. Recently, it has been pointed out that very weakly
unstable dark matter bosons in the mass range below 1 MeV (bosonic
super-WIMPs) can be detected via its absorption by atomic shells
\cite{Pospelov}. The deposited energy is dominated by the dark matter
rest mass, and leads to the $\gamma$-like mono-energetic energy
release. Several experimental analyses exploiting such possibility
have been performed~\cite{Cogent,CDMS}. (Note however, that the suggestion  
that the annual modulation signal observed by DAMA and DAMA-Libra
\cite{DAMA1} is caused  by the absorption of super-WIMPs~\cite{DAMA2}
does not hold because inelastic processes do not have velocity
modulation~\cite{Pospelov}.) Future large-scale experiments with
self-shielding capabilities \cite{Xenon100,LUX,DEAP} will be able to
improve the sensitivity to the super-WIMPs by several orders of
magnitude. 

So far the calculations of absorption rate have been done using simple
models of atoms which ignore the relativistic and many-body
effects~\cite{Pospelov}. This way one can relate the 
absorption cross-sections for axial, scalar
and vector dark matter particles to the photoionization
cross-sections. It is clear however, that atomic corrections 
can be non-negligible, and dedicated atomic calculations of super-WIMP
absorption rates by atoms are required. 

In the present work we perform {\it ab initio} calculations of the
ionization cross section of atoms by pseudoscalar particles using the
relativistic Hartree-Fock (RHF) method. For concision, we call such a
pseudoscalar particle "an axion", stressing that the mass$-$coupling
range considered in this paper does not correspond to a QCD
axion. Our calculations are not linked to the photoionization and can
be directly used for the analysis of the experimental data. We
consider in particular the ionization of several atoms such as argon,
germanium and xenon since these elements are used in the detectors
(see, e.g.~\cite{Avignone,Andriamonje}).  

We have fitted the RHF results by an analytical formula.
The formula works very well when axion energy is
sufficiently large to excite an electron from the $1s$, $2s$ or $2p$
states in atomic core.
%!!!
 We checked that the discrepancy between the
 Hartree-Fock and analytical results
is at the level of a few per cent for axion energy between 10 and 50
keV and nuclear charge $18<Z<60$.
 The formula can be used for any multi-electron atom and wide range of axion
 energies.
The results of this work are to be used in the experimental analyses 
searching for dark matter using underground detectors.

\section{Theory}

The Hamiltonian for the pseudoscalar particle $a$ interacting with
electrons can be written as \cite{Pospelov}
\begin{equation}
  \hat H_a = \frac{\partial_\mu a}{f_a}  \bar \psi \gamma^\mu
  \gamma^5 \psi,
\label{eq:ha}
\end{equation}
where energy scale parameter $f_a$ parametrizes the strength of the
interaction; $\psi$ is electron Dirac field. We do not include the 
direct interaction of axion with photons $a F_{\mu\nu}^2$ on account 
of strong constraints from the $\gamma$-ray backgrounds \cite{Pospelov}. 
It is convenient to present the total cross section of the atom
ionization by axion in a form where 
the part which needs numerical analysis is presented as a
dimensionless factor. Therefore we write
\begin{equation}
  \sigma_a(\epsilon_a) = \left(\frac{\epsilon_0}{f_a}\right)^2 \frac{c}{v}
  K(\epsilon_a) a_0^2 ,
\label{eq:csa}
\end{equation}
where $\epsilon_0$ is an energy scale (in our calculations $\epsilon_0$=1
a.u. = 27.21 eV, but it can also be any other energy unit), $c$ is speed 
of light, $v$ is the axion velocity in the laboratory frame, 
$a_0 = 0.52918 \times 10^{-8}$ cm is Borh radius, 
$\epsilon_a$ is axion energy and 
%??? \epsilon_a
$K(\epsilon_a)$ is dimensionless factor given by 
\begin{equation}
  K(\epsilon_a) = \pi
  \left(\frac{\epsilon_a}{\epsilon_0}\right)^2\epsilon_a \sum_c 
  R_{c\epsilon}^2 (2j_c+1).
\label{eq:k}
\end{equation}

Here summation goes over all electron states $c$ in the atomic core,
$2j_c+1$ is the occupation number of the subshell $c$,
%!!!
$j_c$ is the core electron angular momentum, $R_{c\epsilon}$
is the radial integral for the Hamiltonian (\ref{eq:ha}) with two
single-electron wave functions. Initial electron state is the state
$c$ in atomic core while final electron state is the state in
continuum with the energy
 $\epsilon = \epsilon_a + \epsilon_c$.
The integral is non-zero only between states of the same total
angular momentum $j$ but of the opposite parity.
The integral can be written as
\begin{equation}
 R_{c\epsilon} = \frac{1}{\epsilon_a}\int (f_c(r)g_{\epsilon}(r) -
 g_c(r)f_{\epsilon}(r)) dr,
\label{eq:re1}
\end{equation}
where $f$ and $g$ are upper and lower radial components of the
electron wave function in the spherically-symmetric case:
\begin{equation}
  \psi(\mathbf r) = \frac{1}{r} \left( \begin{array}{r} f(r)
    \Omega_{jlm}(\mathbf{n}) \\ i\alpha
    g(r)\tilde{\Omega}_{jlm}(\mathbf{n}) \end{array} \right).
\label{eq:wf}
\end{equation}
Here $\alpha$ is the fine structure constant ($\alpha \approx 1/137.036$).
The states in the continuum are normalized to the $\delta$-function of
the energy. 
%???(in units $\omega_0$).
Notice that we use realistic estimates of velocities of dark matter 
particles, $v/c \sim 10^{-3}$, which allows to good accuracy retain
only the time derivative term in Eq. (\ref{eq:ha}), and the only
velocity dependence is then  $1/v$ in Eq. (\ref{eq:csa}). 

We use relativistic Hartree-Fock (RHF) method to calculate $K$ using
(\ref{eq:k}) and (\ref{eq:re1}). We will focus on the axion
%!!! 50 KeV
mass window 10 keV$<\epsilon_a<50 $ keV,
which is in the maximal sensitivity range for the
dark matter search experiments. Above these energies, the 
axion-induced Compton like scattering $ae\to e\gamma$ 
may dominate over direct axio-absorption \cite{Pospelov}. As stated
before, we consider 
cold non-relativistic axions so that we can always assume that 
$\epsilon_a = m_ac^2$, where $m_a$ is axion mass.

In principle, formula (\ref{eq:k}) can be used at any axion
energy. However, at low energy, when only outer atomic electrons can
be excited, RHF approximation might be insufficiently accurate due to
many-body and environmental effects not included into the RHF
potential. On the other hand, when ionization is dominated by
excitation from $K$ and $L$ atomic shells these effect can be neglected.
%!!!
As known, the  correlation corrections to RHF results for inner electrons
 $1s,2s,2p$ decrease as $1/Z$ where $Z$ is the nuclear charge.
Moreover, RHF often gives reasonable results for $1s$ even for $Z=2$.
As a next step, the cross section can be approximated by an analytical formula
 which fits RHF results.
%!!! Such formulae exist in scientific literature for
%!!!the case of photoionization. Below we adopt them for the case of axion.
 The advantage of having such formulae is obvious: the results
can be easily obtained for any atom. 

For the non-relativistic operator (but relativistic wave functions)
the radial integral (\ref{eq:re1}) becomes 
(in atomic units)
\begin{equation}
 R_{c\epsilon} = \int (f_c(r)f_{\epsilon}(r) + \alpha^2
 g_c(r)g_{\epsilon}(r)) rdr.
\label{eq:rce}
\end{equation}
This is the same form of the radial integral which appear in the
photoionization cross section. In this case the expressions for the
ionization cross sections by axion and photon are very similar and
differ by a factor only. It was found in Ref.~\cite{Pospelov} that
the cross section of atom ionization by non-relativistic axions can be
expressed via the photoionization cross section $\sigma_\gamma$, 
\begin{equation}
  \frac{\sigma_av}{\sigma_\gamma(\hbar \omega=\epsilon_a)c} \approx
    \frac{3\epsilon_a^2}{4\pi\alpha f_a^2}.
\label{eq:r}
\end{equation}
%where $m_a$ is an axion mass. 
Using expressions (\ref{eq:csa}) and  (\ref{eq:r}) we can present
photoionization cross section in a form 
\begin{equation}
  \sigma(\epsilon_{\gamma}) = \frac{4}{3}\pi\alpha 
\left(\frac{\epsilon_0}{\epsilon_{\gamma}}\right)^2
K(\epsilon_{\gamma}) a_0^2
\label{eq:sg}
\end{equation}
which uses the same dimensionless function of energy $K(\epsilon)$
(\ref{eq:k}) as the axion cross section. This gives us an opportunity
to check numerical calculations using experimental data for the
photoionization. We have done this for the cases of krypton and xenon
using the data from Ref.~\cite{Wuilleumier}. We found that numerical
and experimental results agree within a few per cent accuracy with an
exception of the near the threshold ionization where the difference is
a little higher than 10\%.

\begin{table} % [h]
\caption{Hartree-Fock energies of the core states of Na, Ar, Ge, I and
  Xe (atomic units, 1 a.u.= 27.21 eV).} 
\label{tab:en}
\begin{tabular}{cccccc}
\hline \hline
Atom  & Na    &  Ar  & Ge  &  I  &  Xe  \\
$Z$   & 11    &  18  & 32  & 53  &  54  \\
\hline
$1s_{1/2}$ & -40.54 &  -119.1 &  -411.0 &  -1225. & -1277. \\ 
$2s_{1/2}$ & -2.805 &  -12.41 &  -53.45 &  -193.0 & -202.4 \\
$2p_{1/2}$ & -1.522 &  -9.631 &  -47.33 &  -180.5 & -189.6 \\
$2p_{3/2}$ & -1.514 &  -9.547 &  -46.14 &  -169.5 & -177.7 \\
$3s_{1/2}$ & -0.1823 & -1.286 &  -7.409 &  -40.52 & -43.01 \\
$3p_{1/2}$ &         & -0.5953 & -5.324 &  -35.34 & -37.66 \\
$3p_{3/2}$ &         & -0.5878 & -5.157 &  -33.21 & -35.32 \\
$3d_{3/2}$ &         &         & -1.616 &  -24.19 & -26.02 \\
$3d_{5/2}$ &         &         & -1.591 &  -23.75 & -25.53 \\
$4s_{1/2}$ &         &         & -0.5687 &  -7.759 & -8.430 \\
$4p_{1/2}$ &         &         & -0.2821 &  -5.868 & -6.452 \\
$4p_{3/2}$ &         &         & -0.2730 &  -5.450 & -5.982 \\
$4d_{3/2}$ &         &         &         &  -2.341 & -2.711 \\
$4d_{5/2}$ &         &         &         &  -2.274 & -2.633 \\
$5s_{1/2}$ &         &         &         & -0.8762 & -1.010 \\
$5p_{1/2}$ &         &         &         & -0.4341 & -0.4925 \\
$5p_{3/2}$ &         &         &         & -0.3903 & -0.4398 \\
\hline \hline
\end{tabular}
\end{table}

Expression (\ref{eq:r}) helps us to find an analytical formula for
the atomic ionization by axion absorption. 
Analytical expressions for photoionization in Coulomb field can be
found in many quantum mechanics textbooks (see,
e.g. \cite{Berestetsky,Bethe}). Using the textbook expressions for
photoionization from $1s$, $2s$ and $2p$ Coulomb states and the expression
 (\ref{eq:r}) we firstly obtain the results for a single electron atom.
Then we use additional  parameters to fit the numerical results of the
 relativistic Hartree-Fock calculations. This way we arrive to the formulae
which describe ionization of multi-electron atoms by absorption of axion:
%!!! end
\begin{eqnarray}
K_{total} &=& K_{1s} + K_{2s} + K_{2p}, \label{eq:kt} \\
K_{1s} &=&
f_1(Z,\epsilon_a+\epsilon_{1s})\frac{384\pi\epsilon_{1s}^4}{(\epsilon_0 Z
  \epsilon_a)^2} \frac{e^{-4\nu_1{\rm arccot}\nu_1}}
{1-e^{-2\pi\nu_1}}, \label{eq:k1s} \\
K_{2s} &=& f_2(Z,\epsilon_a+\epsilon_{2s})\frac{6144\pi
  e_2^3}{\epsilon_0 \epsilon_a^2} 
\left(1+3\frac{e_2}{\epsilon_a}\right) \nonumber \\
&&\times \frac{e^{-4\nu_2{\rm arccot}(\nu_2/2)}}
{1-e^{-2\pi\nu_2}},  \label{eq:k2s} \\
K_{2p} &=& f_2(Z,\epsilon_a+\epsilon_{2p})\frac{12288\pi
  e_3^4}{\epsilon_0 \epsilon_a^3} 
\left(3+8\frac{e_3}{\epsilon_a}\right) \nonumber \\
&&\times \frac{e^{-4\nu_3{\rm arccot}(\nu_3/2)}}{1-e^{-2\pi\nu_3}}, 
 \label{eq:k2p}
\end{eqnarray}
where $\alpha$ is the fine structure constant, $Z$ is nuclear
charge, $\epsilon_a$ is axion energy, $e_2=|\epsilon_{2s}|$,
$e_3=|\epsilon_{2p}|$,  $\nu_1 =
\sqrt{-\epsilon_{1s}/(\epsilon_{1s}+\epsilon_a)}$,  $\nu_2 =
2\sqrt{-\epsilon_{2s}/(\epsilon_{2s}+\epsilon_a)}$,  $\nu_3 =
2\sqrt{-\epsilon_{2p}/(\epsilon_{2p}+\epsilon_a)}$.  
Here $\epsilon_{1s}$, $\epsilon_{2s}$ and  $\epsilon_{2p}$ are the
Hartree-Fock energies of the core states. Hartree-Fock energies of all
core states of Na, Ar, Ge, I and Xe are presented in
Table~\ref{tab:en}. For other atoms extrapolation formulas can be used:
\begin{eqnarray}
%  \frac{\epsilon_{1s}}{\omega_0}(Z) = -(Z^2-7.49Z+43.39)/2
  \frac{\epsilon_{1s}}{\epsilon_0}(Z) &=& -\frac{Z^2-7.49Z+43.39}{2},
\label{eq:1s} \\
  \frac{\epsilon_{2s}}{\epsilon_0}(Z) &=& -0.000753 Z^3-0.028306 Z^2 \nonumber \\
&&-0.066954 Z \ +2.359052, \label{eq:2s} \\
  \frac{\epsilon_{2p}}{\epsilon_0}(Z) &=& -0.000739 Z^3-0.027996 Z^2 \nonumber \\
&&+0.128526 Z \ +1.435129. \label{eq:2p} 
\end{eqnarray}
The functions $f_1(Z)$ and $f_2(Z)$ in
(\ref{eq:k1s},\ref{eq:k2s},\ref{eq:k2p}) are scaling functions:
%???  \epsilon/\epsilon_0
\begin{eqnarray}
%         f_1(Z,\epsilon) &=& (-0.00214Z^2 + 0.2673Z -
%         22.66)\times 10^{-5}\epsilon/\epsilon_0 \nonumber \\
%         &+&  0.0004736Z^2 - 0.05596Z + 2.454 \label{eq:f1} \\
%         f_2(Z) &=&   0.000705Z^2 - 0.087Z + 3.18.
         f_1(Z,\epsilon) &=& (5.368 \times 10^{-7}Z -
         1.17\times 10^{-4})\epsilon/\epsilon_0 \nonumber \\
         &&  - 0.012Z + 1.598 \label{eq:f1} \\
         f_2(Z,\epsilon) &=& (-1.33 \times 10^{-6}Z +
         1.17\times 10^{-4})\epsilon/\epsilon_0 \nonumber \\
         &&  - 0.0156Z + 1.15 \label{eq:f2} 
\end{eqnarray}
They are chosen to fit the results of Hartree-Fock calculations for
axion energies between 10 and 50 keV and nuclear charge $18<Z<60$.
 We stress once more that not
only scaling functions but all the formulae
(\ref{eq:k1s}-\ref{eq:f2})  
were obtained by fitting the Hartree-Fock calculations. 
All atomic shells are included into the RHF calculations. Therefore,
the fit with $K_{total} = K_{1s}+K_{2s}+K_{2p}$ includes the higher
shells as well. As it will be demonstrated in the next section, the
analytical and Hartree-Fock results agree within few per cent.
%!!! This means that analytical formulae corresponds to the Hartree-Fock
%!!! approximation rather than Coulomb approximation, which is usually
%!!! used to obtain analytical formulae for atom photoionization.

\section{Results}

\begin{figure}
\centering
\epsfig{figure=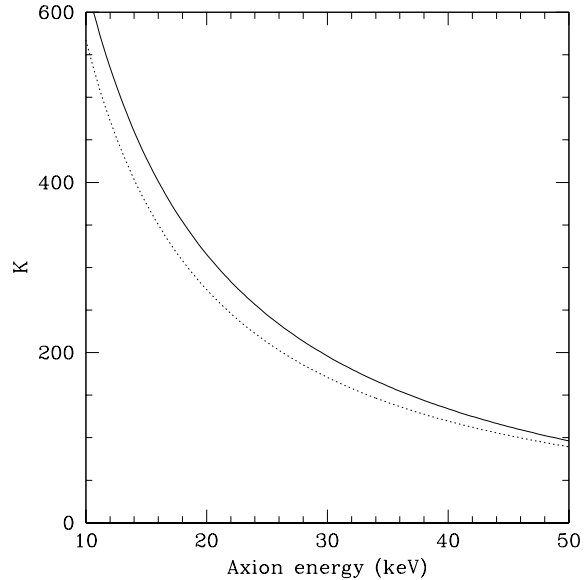,scale=0.40}
\caption{Dimensionless factor $K$ (see formula (\ref{eq:k}) in the
  ionization cross sections of Ar by axion. Solid line -
  result of Hartree-Fock calculations, dotted line - 
formula~(\ref{eq:kt}).}  
\label{fig:ar}
\end{figure}

\begin{figure}
\centering
\epsfig{figure=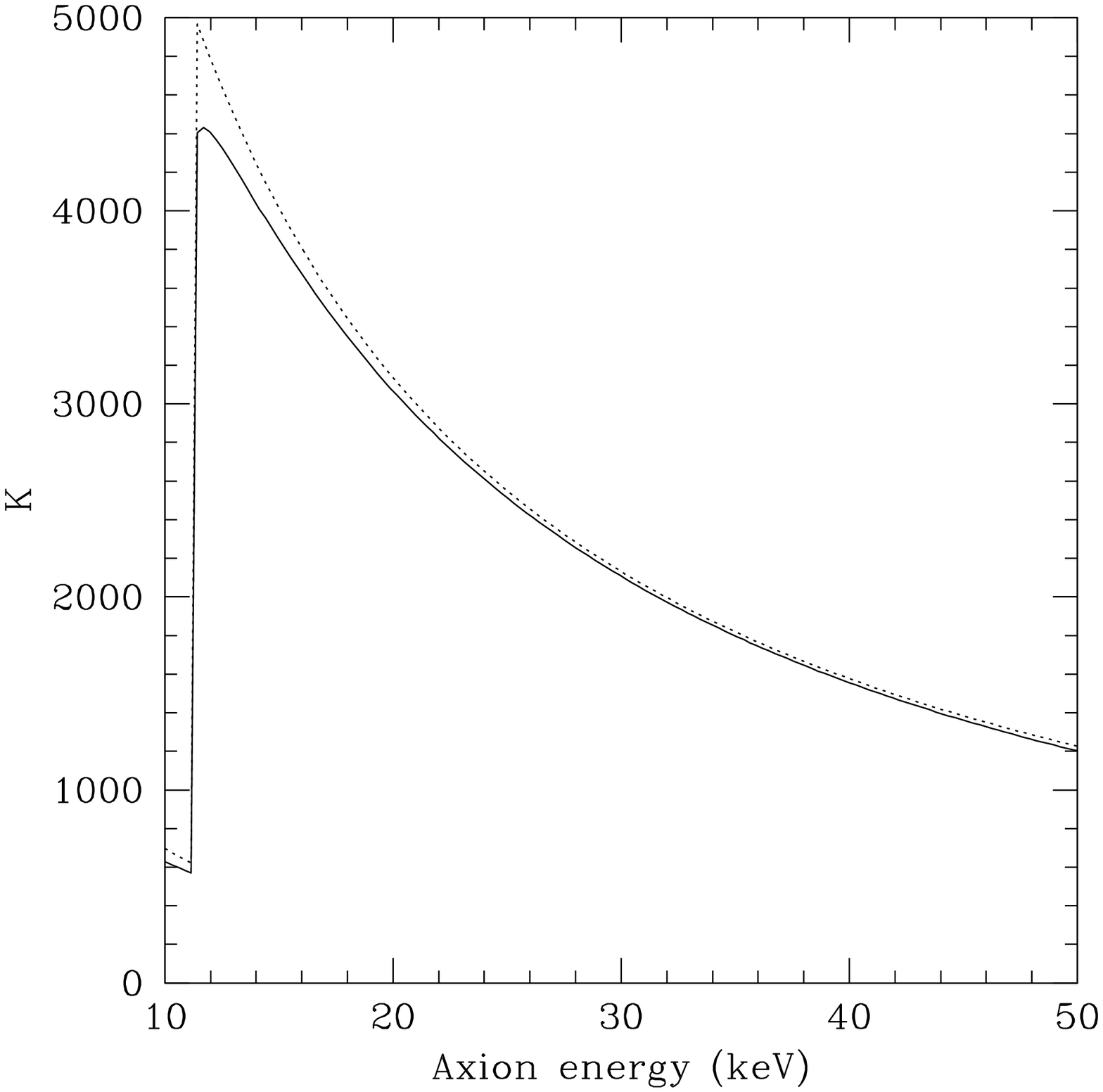,scale=0.40}
\caption{As on Fig.~\ref{fig:ar} but for Ge.}
\label{fig:ge}
\end{figure}

\begin{figure}
\centering
\epsfig{figure=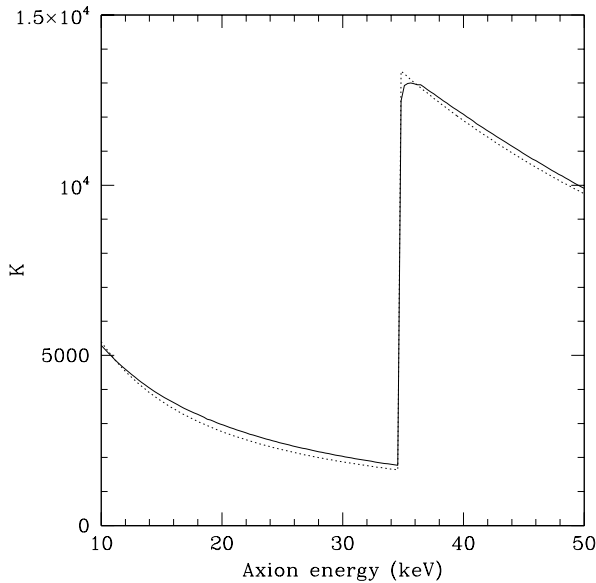,scale=0.40}
\caption{As on Fig.~\ref{fig:ar} but for Xe.}
\label{fig:xe}
\end{figure}

Figures \ref{fig:ar},\ref{fig:ge},\ref{fig:xe} show the results of 
relativistic Hartree-Fock and analytical calculations of the
dimensionless factor $K$ which stands in the expression for the cross
section of atom ionization by axion (see formula
(\ref{eq:csa})). We use the RHF computer code and formulae (\ref{eq:k})
and (\ref{eq:re1}) for the calculations. Many body and relativistic
effects beyond the RHF method are ignored and the final electron state
in the continuum is calculated in the same potential as initial core
state. The accuracy of this approximation is few percents due to
dominating contribution from the inner-most core states $1s$, $2s$ and
$2p$. For these states the many-body effects are small due to strong
nuclear field.  

Calculations using analytical formula
(\ref{eq:kt},\ref{eq:k1s},\ref{eq:k2s},\ref{eq:k2p}) are shown
on Figures \ref{fig:ar},\ref{fig:ge},\ref{fig:xe} as dotted lines.
The agreement between RHF and analytical results is very
good for all three atoms. It is on the order of a few per cent when an electron
from the $1s$ is excited and better than 20\% when only $2s$ and
higher states contribute to the cross section.
We expect similar trend for other atoms.
This means that at present stage the analytical formulae 
(\ref{eq:kt},\ref{eq:k1s},\ref{eq:k2s},\ref{eq:k2p}) are sufficiently
accurate and there is no need for more sophisticated atomic
calculations. 

The results show an obvious advantage of using heavy elements in the
detectors given that axion energy is sufficiently large to excite an
electron from an inner-most subshell. For example, for a solar axion
of $\epsilon_a$=14.4 keV~\cite{solar}, $K=450$ for the case of Ar
and $K \approx 4000$ for the case of Ge and Xe. Current experimental 
sensitivity to the $f_a$ parameter is in the range of $10^8-10^9$ GeV
\cite{Cogent,CDMS}, 
but is poised to be improved in the very near future. 

In this paper we have examined the case of atomic ionization by the
pseudoscalar  super-weakly interacting dark matter, but a similar case 
can be made for other types of super-WIMPs with arbitrary integer
spin, and coupled to electrons via vector, axial-vector, tensor etc 
couplings. Another obvious use of the RHF calculations relevant to the
underground searches of exotic particles, is the generalization to the
detection of nearly massless particles emitted by the solar interrior
with typical energies of a few keV. Both applications will be
considered in a forthcoming publication.

\acknowledgements
The authors are grateful to M. Kuchiev and J. Berengut for useful
discussions.
The work was supported in part by the Australian Research Council.

\end{document}